\documentclass[twocolumn, prl, amssymb, superscriptaddress, aps, showpacs,preprintnumbers,
amsmath,floatfix]{revtex4}

\newcommand{\be}{\begin{equation}}
\newcommand{\ee}{  \end{equation}}
\newcommand{\ba}{\begin{eqnarray}}
\newcommand{\ea}{  \end{eqnarray}}

\usepackage{epstopdf}
\usepackage{graphics}
\usepackage{graphicx}
\usepackage{dcolumn}
\usepackage{bm}
\usepackage{longtable}
\usepackage{epsfig}
\usepackage{times}
\usepackage{url}
\usepackage{color}
\usepackage{subfigure}

\begin{document}
%%%%%%%%%%%%%%%%%%%%%%%%%%%%%%%%%%%%%%%%%%%%%%%%%%%%%%%%%%%%%%%%%%%%%%%%%%%%%%%%

\title{Laser-Matter Interaction: Classical Regime versus Quantum Regime}
\author{Adriana \surname{P\'alffy}}
\email{palffy@mpi-hd.mpg.de}
\affiliation{Max-Planck-Institut f\"ur Kernphysik, Saupfercheckweg 1, D-69117 Heidelberg, Germany}

\author{Paul-Gerhard \surname{Reinhard}}

\affiliation{Institute for Theoretical Physics II,
University of Erlangen-Nuremberg,
Staudtstr. 7, 91058 Erlangen, Germany}

\author{Hans A. \surname{Weidenm\"uller}}
\email{haw@mpi-hd.mpg.de}
\affiliation{Max-Planck-Institut f\"ur Kernphysik, Saupfercheckweg 1, D-69117 Heidelberg, Germany}

\date{\today}
%%%%%%%%%%%%%%%%%%%%%%%%%%%%%%%%%%%%%%%%%%%%%%%%%%%%%%%%%%%%%%%%%%%%%%%%%%%%%%%%

%%%%%%%%%%%%%%%%%%%%%%%%%%%%%%%%%%%%%%%%%%%%%%%%%%%%%%%%%%%
\begin{abstract}

Doppler backscattering of optical laser photons on a ``flying mirror'' of
relativistic electrons promises to yield coherent photons with
MeV-range energies. We compare the nuclear interaction of such a laser
pulse with the standard atom-laser interaction. The mean-field
description of atoms must be replaced by a rate equation and the
classical field strength, far too faint in nuclei, by the dipole
transition rate. Significant nuclear excitation occurs for photon
numbers much smaller than typical for atoms. That drastically
reduces the requirements on the experimental realization of a ``flying
mirror''.

\end{abstract}
%%%%%%%%%%%%%%%%%%%%%%%%%%%%%%%%%%%%%%%%%%%%%%%%%%%%%%%%%%%


% 42.50.Ct 	Quantum description of interaction of light and matter; related experiments
% 25.20.-x 	Photonuclear reactions
% 24.60.Dr 	Statistical compound-nucleus reactions 
% 32.80.Rm 	Multiphoton ionization and excitation to highly excited states

%%%%%%%%%%%%%%%%% other possibilities 
% 25.70.Gh 	Compound nucleus 
% 52.38.Ph 	X-ray, γ-ray, and particle generation 
% 21.10.Pc 	Single-particle levels and strength functions 

%%%%%%%%%%%%%%%%%%%%%%%%%%%%%%%%%%%%%%%%%%%%%%%%%%%%%%%%%%%%%%%%%%%%%%%%%%%%%%%%
\maketitle
%%%%%%%%%%%%%%%%%%%%%%%%%%%%%%%%%%%%%%%%%%%%%%%%%%%%%%%%%%%%%%%%%%%%%%%%%%%%%%%%

{\it Introduction.} This paper is triggered by recent experimental,
computational and theoretical advances in the production of
high-energy coherent laser pulses, based on the following
mechanism~\cite{Ein05}. A first intense laser pulse ejects electrons
from a nanometer-thin Carbon foil. The electrons attain relativistic
energies and form a ``flying mirror''. On that mirror, a second laser
pulse is Doppler backscattered~\cite{Esi2009,Kie2009,Mey2009,Mou11,
  Kie13,Bul2013,Mu2013,Li2014}. That increases both the energy $\hbar
\omega_0$ and the energy spread $\sigma$ of the photons in the second
pulse by a factor $2 \sqrt{1 - (v/c)^2}$ where $v$ is the velocity of
the ejected electrons. In principle, photon energies in the MeV range
and beyond can be reached, accompanied by energy spreads in the $50$
keV range and beyond.

Doppler backscattering of photons on a ``flying mirror'' of electrons
has produced coherent photons in the far ultraviolet
regime~\cite{Kie13} but not yet coherent MeV photons. Attaining such
energies apparently requires a further step. The electrons in the
flying mirror must be compressed to a mean density that is close to
condensed-matter values~\cite{Thi2016}. The present work anticipates a
positive solution of that problem. We accordingly consider a coherent
laser pulse with a typical energy $\hbar \omega_0 \approx 5$ MeV per
photon and with a typical energy spread $\sigma$ in the $50$ keV
range.

The prospect of a laser beam with photon energies comparable to
typical nuclear excitation energies raises important questions. What
is the difference between the (well-investigated) laser-atom and the
(novel) laser-nucleus interaction? What conditions follow from that
comparison for the experimental realization of the
Doppler-backscattered laser pulse? In the main body of this Letter we
answer the first of these questions. We compare the interaction
with a medium-weight or heavy nucleus of a pulse as specified above
with that of a standard laser pulse with an atom. The answer to the
second question emerges in the form of a corollary.

Laser-induced multiphoton excitation in atoms and in nuclei are
fundamentally different processes, for two reasons. (i) In atoms, the
Coulomb interaction between electrons has long range and is
repulsive. It can often be accommodated in terms of a mean field. In
lowest order, electrons are then described as a system of independent
particles \cite{Riv09,Ton10}. In nuclei, the residual nucleon-nucleon interaction beyond
the mean field has short range, is strong, and drives the nucleus
towards equilibrium. Reactions induced on medium-weight and heavy
nuclei at excitation energies in the $10$ MeV range and beyond are,
therefore, described by rate equations~\cite{Dur00,Wei07}, a regime which is also exploited in atomic/molecular systems, however
at comparatively larger energies \cite{Dit96}. (ii) For
multiphoton excitation in atoms, the use of the classical
approximation for the electric field strength of the laser is standard
and defines the classical regime \cite{Fai87}. Scaling of that field strength and
of the dipole operator shows that this approach does not lead to
significant excitation in nuclei. In contradistinction, scaling shows
that the dipole transition rate is much enhanced in nuclei. That fact,
combined with the use of rate equations for equilibration, defines the
quantum regime and yields a satisfactory theoretical framework for
laser-nucleus reactions. The number of photons per pulse required for
multiple nuclear excitation via rate processes is orders of magnitude
smaller than the number needed within the classical regime for
multiple atomic excitation \cite{Gav92}. That fact drastically reduces the
requirements for a successful experimental realization of the ``flying
mirror''.

In what follows we address both the quantum-optics community and the
nuclear-physics community. The paper is written in that spirit. 

{\it Coupling to the Electromagnetic Field.} With S denoting the
quantum system (atom or nucleus) and F the quantized electromagnetic
field, the total Hamiltonian is
\be
H = H_{\rm S} + H_{\rm F} + H_{\rm int} = H_0 + H_{\rm int} \ .
\label{1}
\ee
For atoms (nuclei), the relevant photon energy is in the eV range (the
MeV range, respectively). In either case the product of wave number
$k$ and atomic (nuclear) radius $R$ obeys $k R \ll 1$. That justifies
the use of the dipole approximation. The interaction part of $H$ is
then $H_{\rm int} = - e \vec{q} \vec{E}(\vec{r})$. Here $\vec{E}$
denotes the operator of the free electric field strength, taken at the
position $\vec{r}$ of the atomic nucleus. In the atomic case,
$\vec{q}$ denotes the sum of the position operators of the electrons
relative to the atomic nucleus. In the nuclear case, $\vec{q}$ is
proportional to the difference of the centers of mass of neutrons and
protons~\cite{Rin80}. The inner (direct) product of two vectors ${\vec
  a}$ and $\vec{b}$ is written as $\vec{a} \vec{b}$ (as $\vec{a}
\rangle \langle \vec{b}$, respectively).  With $t$ denoting the time,
we use the interaction representation where ${\cal H}_{\rm int} = \exp
\{ i H_0 t / \hbar \} \ H_{\rm int} \ \exp \{ - i H_0 t / \hbar \}$
and correspondingly for $\vec{q}(t)$ and $\vec{E}(\vec{r}, t)$. Then
\be
{\cal H}_{\rm int} = - e \vec{q}(t) \vec{E}(\vec{r}, t) \ .
\label{4}
\ee
Needless to say, Eq.~(\ref{4}) applies only while the laser pulse
lasts. Transients due to onset and termination of the laser pulse are
neglected.

The electric field strength $\vec{E}$ is expanded in a set of
orthonormal modes, defined~\cite{Gla07} in a large but finite cubic
normalization volume of side length $L$ with periodic boundary
conditions. The modes are polarized plane waves $L^{- 3/2}
\vec{e}_\lambda \exp \{ i {\vec k} {\vec r} \}$ with discrete wave
vectors ${\vec k}$. The two polarization vectors $\vec{e}_\lambda$
with $\lambda = 1, 2$ are orthogonal upon each other and upon
$\vec{k}$.  For brevity we use a joint index $k = ({\vec k}, \lambda)$
for the associated creation and annihilation operators $a^{\dag}_k$
and $a^{}_k$ which obey $[a^{}_{k}, a^{}_{k'}] = 0 = [a^\dag_{k},
  a^\dag_{k'}]$, $[a^{}_{k}, a^{\dag}_{k'}] = \delta_{k k'}$. With
$\omega_k = c |\vec{k}|$ we then have $H_{\rm F} = \frac{1}{2} \sum_k
\hbar \omega_k ( a^\dag_k a^{}_k + a^{}_k a^\dag_k )$. The expansion
for the electric field strength reads~\cite{Gla07}
\be
{\vec E}({\vec r}, t) = i \sum_k \ \sqrt{\frac{\hbar \omega_k}{2}}
\frac{1}{L^{3/2}} {\vec e}_{\lambda} \bigg[ \exp \{ i {\vec k}
{\vec r} - i \omega_k t \} a_k - h. c. \bigg] \ .
\label{8}
\ee
We use a basis of coherent states. For fixed $k$ and arbitrary complex
$\alpha_k$ the normalized coherent state $| \alpha_k \rangle$ obeys
$a_k | \alpha_k \rangle = \alpha_k | \alpha_k \rangle$. The
expectation value $n_k$ of the number of photons in mode $k$ and the
total number $N$ of photons in the pulse are
\be
n_k = \langle \alpha_k | a^\dag_k a_k | \alpha_k \rangle = | \alpha_k
|^2 \ , \ N = \sum_k n_k \ .
\label{10}
\ee

{\it Description of the backscattered laser pulse.} We use three
assumptions. First we assume~\cite{Gla07} that the density matrix
$\rho(\{ \alpha_k \})$ of the laser pulse is stationary in time.
Stationarity is plausible for pulse lengths considered here that are
about two orders of magnitude larger than the wave length.
Stationarity is equivalent to phase-averaging over the
amplitudes $\alpha_k$ and implies ${\rm Tr} [ \rho a_k ] = 0$ and
${\rm Tr} [ \rho a^\dag_k a_{k'} ] = n_k \delta_{k k'}$~\cite{Gla07}.
For stationary fields the expectation value of the field strength,
therefore, vanishes, and the field strength must be defined via the
square root of the intensity. We write ${\vec E} = {\vec E}^+ + {\vec
  E}^-$ where ${\vec E}^+$ (${\vec E}^-$) contains the annihilation
(the creation) operators, respectively. Then
\ba
&& {\rm Tr} [ \rho \vec{E}^-(\vec{r}\,^\prime, t') \rangle \langle
\vec{E}^+(\vec{r}, t) ] = \nonumber \\
&& \ \ \sum_{k} \frac{\hbar \omega_k}{2 L^3} n_k {\vec e}_{\lambda}
\rangle \langle {\vec e}_{\lambda} \exp \{ i {\vec k} ({\vec r} -
\vec{r}\,^\prime) - i \omega_k ( t - t') \} \ .
\label{11}
\ea
Second we assume that the laser is sufficiently monochromatic. We may
then replace in Eq.~(\ref{11}) $\omega_k$ by the mean frequency
$\omega_0$, $\vec{k}$ by $\vec{k}_0$ and $\vec{e}_\lambda$ by
$\vec{e}_{\lambda_0}$, respectively. We use Eq.~(\ref{10}). Then the
right-hand side of Eq.~(\ref{11}) factorizes and can be written as the
product $E^-_{\rm cl}(\vec{r}\,^\prime, t') \rangle \langle E^+_{\rm
  cl}(\vec{r}, t)$ of the components
\be
{\vec{E}^\pm_{\rm cl}(\vec{r}, t}) = \sqrt{\frac{N \hbar \omega_0}{2 L^3}} 
\vec{e}_{\lambda_0}  \exp \{ \pm (i {\vec k}_0 {\vec r} - i \omega_0 t ) \}
\label{12}
\ee
of the classical electric field strength. Third, we assume that the
``flying mirror'' is not exactly planar. Then the backscattered pulse
is not completely collinear but has a finite aperture.

{\it Classical Regime.} The expression
\be
{\cal H}_{\rm int}(\vec{r}, t) = - e \vec{q}(t) \vec{E}_{\rm cl}(\vec{r},
t)
\label{15}
\ee
for the interaction Hamiltonian defines the classical regime. We have
to bear in mind, however, several restrictions. First, Eqs.~(\ref{12})
and (\ref{15}) apply only in a restricted domain of space and time as
defined by the coherence properties of the laser pulse. For a
laser-induced nuclear reaction, quenching of the original laser pulse
in the direction of propagation by Doppler backscattering results in a
pulse that has approximately the shape of a circular disk with lateral
radius $r \approx$ several $\mu$m (the value for the pulse prior to
backscattering) and length $l \approx \hbar c / \sigma \approx 100$
wave lengths so that $l \ll r$. Accordingly, the lateral spread in momentum space is 
much smaller than the value $k_0$.
The second restriction
emerges when we consider terms of higher order than the first in
${\cal H}_{\rm int}$. The transition probability contains a sum of
traces of $\rho (\vec{E}^+)^m (\vec{E}^-)^n$, with integer $m,
n$. Stationarity of $\rho(\{ \alpha_k \})$ implies that only terms
with $m = n$ differ from zero. Use of the classical
Hamiltonian~(\ref{15}) is possible only if for all $n$, ${\rm Tr} [
\rho (\vec{E}^+)^n (\vec{E}^-)^n ]$ factorizes in the same manner as
does the right-hand side of Eq.~(\ref{11}). Then the field is fully
coherent (the $n^{\rm th}$-order correlation functions factorize for
all $n$).  That is the case for coherent laser beams where the field
is generated by an ``essentially classical source''~\cite{Gla07}. The
third restriction is the most severe one. It arises because (as we
have just noted) in the quantum approach the operators $\vec{E}^+$ and
$\vec{E}^-$ always contribute with equal powers to the transition
probability. The corresponding property does not hold for the
classical Hamiltonian where the transition probability does contain
nonvanishing terms proportional to $(\vec{E}^+_{\rm cl})^m
(\vec{E}^-_{\rm cl})^n$ with $m \neq n$. Dipole transitions connect
only states of opposite parity. That constraint somewhat reduces the
number of combinations $(\vec{E}^+_{\rm cl})^m (\vec{E}^-_{\rm cl})^n$
with $m \neq n$. Nevertheless the semiclassical
approximation~(\ref{15}) is at best semiquantitatively correct.

{\it Single-Photon Process.} The quantum regime is here defined by the
use of rate equations. We consider dipole-induced one-step photon
absorption in the quantum system S from an initial state $| i \rangle$
to a final state $| f \rangle$. The process involves only
$\vec{E}^+(\vec{r}, t)$. The trace of the square of the transition
matrix element then carries the factor ${\rm Tr} [\rho \vec{E}^-
\rangle \langle \vec{E}^+ ]$. As shown above, the transition is
induced by the monochromatic electric field of Eq.~(\ref{12}). The
calculation of the rate is, except for the factor $\sqrt{N}$,
identical to that for a single photon of frequency $\omega_0$, and is
standard~\cite{Scu97}. First-order time-dependent perturbation theory
yields the transition amplitude $b_{i \to f}(t)$. For the square of
that expression, we use the long-wavelength limit. The
approximation~(\ref{12}) holds only within a finite domain of space
and time, and it is necessary to sum over the finite width of the
photon $\vec{k}$, including the finite width of the aperture. That is
done using~\cite{Gla07} $\sum_k \to (L^3 / (2 \pi)^3) \int {\rm d}^3
k$. The last step involves the limit $L \to \infty$ and removes the
normalization volume $L^3$. For an on-shell transition ($E_f - E_i =
\hbar \omega_0$) the dipole absorption rate ${\cal R}_{\rm dip} = (1 /
t) \sum_k | b_{i \to f}(t) |^2$ is then given by
\be
{\cal R}_{\rm dip} = \frac{e^2}{\hbar c} \ N \ \frac{2}{\pi}
\omega_0 k^2_0 \ \sum | \langle f | (\vec{q} \ \vec{e}_{\lambda_0}) | i
\rangle |^2 \ .
\label{26}
\ee
We have used spherical polar coordinates. The sum indicates that the
integral over the solid angle, i.e., the aperture of the pulse, the
average over initial spin directions, and the sums over polarization
directions and final spin directions have yet to be carried out. 
The aperture of the pulse is determined by the backscattering process. 
We assume that the experiment can be 
driven in such a way that the aperture becomes quite small and only weakly dependent on $k_0$. Then
the solid-angle integral, though strongly suppressing the rate
in comparison with its full solid-angle value, does not remove
the $k_0^3$-dependence of expression (\ref{26}). That dependence differs
from the one of a completely collinear optical laser~\cite{Lou2009}  where
the rate is linear in $k_0$ because 
one considers only the integral over the spread of the photon frequency $\omega$.

 The
dipole width is $\Gamma_{\rm dip} = \hbar {\cal R}_{\rm dip}$. Without
the factor $N$, the dipole width in nuclei is of the order $100$ eV -
$1$ keV. For $N = 10^3 - 10^4$ photons in the pulse, the nuclear
dipole width is boosted into the MeV domain. Such values were used in
the calculations of Refs.~\cite{Pal14, Pal15}. 

{\it Multiphoton Process.} An exact treatment of a laser-induced
multiphoton process would use time-dependent perturbation theory for
${\cal H}_{\rm int}(t)$ and the eigenstates of $H_{\rm S}$ as
intermediate states. Such an exact treatment may, however, be
impractical or even beyond reach in interacting many-body systems. Two
approximation schemes may be used: a mean field-approach combined with
the classical approximation for the electric field (i), and rate
equations (iii).

(i) Multiphoton Process in Atoms. The interaction Hamiltonian ${\cal
  H}_{\rm int}$ is a (sum of) single-particle operator(s). That fact
and the long range of the Coulomb interaction suggests using a
mean-field approximation.  Theoretical work on multiphoton processes
in laser-atom interactions is, therefore, often based upon a
combination of the independent-particle picture and the classical
approximation~(\ref{15}) for ${\cal H}_{\rm int}$. Each electron moves
independently under the influence of the classical electric field
strength $\vec{E}_{\rm cl}$. The field strength~(\ref{12}) is usually
determined phenomenologically.  Empirically, the product $e R
|\vec{E}_{\rm cl}|$ ranges from meV values to maximum values in the eV
range for very strong lasers. Such strong fields deform the nuclear
Coulomb potential, and electrons may be set free by tunnel
ionization. For a laser pulse of frequency $\omega = 1$ eV / $\hbar$,
of length $\hbar c$ / ($10^{- 2}$ eV) and with lateral width of the
order of $10^{- 6}$ m, Poynting's theorem shows that for $e | \vec{E}
|$, values in the range of $(0.001 - 1)$ eV / ($10^{- 8}$ cm) are
attained for photon numbers $N$ in the range $N = 10^6 - 10^{12}$. The
classical approach has important advantages: It offers qualitative and
intuitive insights into the way the laser interacts with atoms. It
allows for further important and useful approximations such as the
Keldysh formalism (see, for instance,
Ref~\cite{Kel65,Mil2006,Pop14}). The time-dependent Schr{\"o}dinger
equation can be solved stepwise numerically while the solution of the
full quantum problem would be much less transparent.

(ii) Nuclei: Scaling. The classical approximation~(\ref{15}) has
little bearing for laser-nucleus interactions because the attainable
classical field strengths are too faint. We show that using a scaling
argument and a numerical simulation.

For the scaling we take an optimistic view, grossly overestimating the
field strength. We use Pointing's theorem $\vec{E}^2_{\rm
  cl} = N \hbar \omega_0 / (2 V_L)$. We assume that all $N$ photons of
the initial laser pulse are coherently Doppler backscattered on the
``flying mirror'' described above. That increases both the mean photon
energy $\hbar \omega_0$ and the energy spread $\sigma$ of the photons
to values of about $5$ MeV and $50$ keV, respectively, i.e., by about
a factor $10^7$. We assume that the volume $V_L$ of the laser pulse is
quenched in the direction of propagation by that same factor while the
lateral width remains unchanged. Then $|\vec{E}_{\rm cl}|$ scales with
the factor $10^7$. For a medium-weight or heavy nucleus, the nuclear
radius is about four orders of magnitude smaller than the radius of
the atom. The interaction Hamiltonian~(\ref{15}) thus scales with a
factor $10^3$. For a strong atomic laser, $e R |\vec{E}_{\rm cl}|$ is
of the order eV. For the nuclear case scaling gives $e R |\vec{E}_{\rm
  cl}| \approx 1$ keV. Typical excitation energies for low-lying
states in medium-weight or heavy nuclei are two orders of magnitude
larger than that figure, and three orders of magnitude larger in light
nuclei. Within the classical approximation, efficient nuclear
excitation seems therefore unlikely. Indeed, with $e R |\vec{E}_{\rm
  cl}| \approx 1$ keV and the duration time of the pulse given by
$\hbar / \sigma$, the nuclear transition amplitude is of order $1$ keV
/ $\sigma \approx 1 / 50$ and the transition probability is of order
$4 \cdot 10^{- 4}$.

\begin{figure}[t]
\centering
\includegraphics[width=0.4\textwidth]{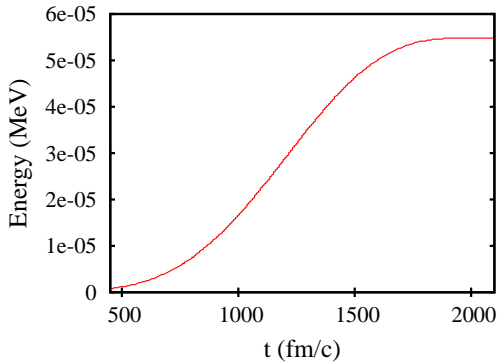}
\caption{\label{fig1}  Energy absorbed by the photon
  field as a function of time.}
\end{figure}

Numerical calculations of photon absorption by $^{16}$O using the
interaction Hamiltonian~(\ref{15}) confirm that scenario. We use a
three-dimensional time-dependent Hartree-Fock code~\cite{sky3D} based
on the Skyrme energy density functional and a cutoff factor to account
for the finite duration of the laser pulse, writing $\hat{\cal H}_{\rm
  int} = {\cal H}_{\rm int} \cos^2 [\pi (t - T_{\rm pulse}) / (2
T_{\rm pulse})]$, where $T_{\rm pulse}$ is the half width of the
pulse. The arguments given above suggest that we use $e | \vec{E}_{\rm
  cl} | = 10^{- 4}$ MeV/fm and $T_{\rm pulse} = 10^{- 19}$ s. That
faint field strength would require extreme precision in the numerical
calculations. We use the fact that we are deeply in the linear regime
and rescale $| \vec{E}_{\rm cl} |$ and $T_{\rm pulse}$, keeping the
fluence, i.e., the product $| \vec{E}_{\rm cl} | T^2_{\rm pulse}$
constant. Our calculations were performed for $| \vec{E}_{\rm cl} | =
0.033$ MeV/fm and $T_{\rm pulse} = 2000$ fm/c. Figure~\ref{fig1} shows
the energy absorbed from the photon field versus time for a photon
energy $\hbar \omega = 10$ MeV. Dividing that number by $10$ MeV gives
the average number of absorbed photons as $5 \cdot 10^{- 6}$. That is
actually an overestimate because we have enhanced $| \vec{E}_{\rm cl}
|$ by a factor $30$. On the other hand, larger absorption rates are
expected for photon energies that are close to the energy of the
dipole resonance and for heavier nuclei. In any case, these results
confirm our estimates and show that the classical
approximation~(\ref{15}) does not yield significant photon absorption
in nuclei.

%%%%%%%%%%%%%%%%%%%%%%%%%%%%%
%%%%%%%%%%%%%%%%%%%%%%%%%%%%%

(iii) Multiphoton Process in Nuclei: Rate Equations. When we describe
a single photon by setting $N = 1$ in Eq.~(\ref{12}), the resulting
classical electric field strength is entirely negligible for exciting nuclei.
Nevertheless, the absorption of single photons with energies in the
MeV range is an important process in nuclei. The apparent discrepancy
is trivially resolved: Single-photon absorption is described not
within the classical approximation~(\ref{15}) but in terms of the
transition rate~(\ref{26}). The rate is proportional to $(\hbar
\omega_0)^3$ and to $R^2$ and, therefore, scales as $(10^7)^3 (10^{-
  4})^2 = 10^{13}$ while we have shown above that the square of the
interaction Hamiltonian scales as $(10^7)^2 (10^{- 4})^2 = 10^6$. The
additional factor $10^7$ appearing in the rate not only explains the
strength of photon absorption in nuclei; The estimate below
Eq.~(\ref{26}) suggests that for $N \gg 1$ the dipole transition rate
also plays a dominant role in laser-induced multiphoton absorption
provided the process is described in terms of rate equations.

When would that be the case? In nuclei the strong residual short-range
interaction (i.e., the nucleon-nucleon interaction beyond the mean
field) drives the system towards statistical equilibrium (the
``compound nucleus''). In a series of two-body collisions, the dipole
mode excited by photon absorption mixes with other modes having
(nearly) the same excitation energy and carrying the same quantum
numbers until all such modes are occupied with equal weight. The
characteristic time scale is that of a nucleon-nucleon collision
induced by the residual interaction \cite{Pin66,Ber83aR}. The collision time $\tau_{\rm
  coll} \approx \lambda / v_{\rm F}$ is the ratio of the nucleon mean
free path $\lambda \approx 4 - 6$ fm and of the Fermi velocity $v_{\rm
  F} = \sqrt{2 E_{\rm F} / m}$ where $m$ is the nucleon mass and
$E_{\rm F} \approx 30$ MeV is the Fermi energy. Thus, $\tau_{\rm coll}
\approx 5 \cdot 10^{- 23}$ s or $\hbar / \tau_{\rm coll} \approx 10$
MeV. We have shown below Eq.~(\ref{26}) that for $N \approx 10^3 -
10^4$ the dipole width attains values in the MeV range, comparable
with $\hbar / \tau_{\rm coll}$. In other words, the nucleus readjusts
to absorption of a single photon within a time that is comparable to
that for the absorption of the next photon.  Therefore, laser-induced
multiphoton absorption in nuclei must take account of relaxation
processes.

In medium-weight and heavy nuclei, the only practicable way of doing
that consists in using rate equations. At excitation energies above
approx. $10$ MeV, these nuclei are so complex that dynamical details
cannot be followed in practice, and a statistical description must be
used. The approach is standard in the theory of preequilibrium
processes induced by projectiles of several $10$ MeV kinetic energy
impinging on nuclei~\cite{Wei07}. Equilibration is described in terms
of the change in time of the average occupation probabilities of
classes of states. Such classes can, for instance, be defined by the
number of particle-hole excitations out of the ground state. The
associated rate equations involve $\hbar / \tau_{\rm coll}$ and
density-of-states factors. Rates for laser-induced photon absorption
and emission derived from ${\cal R}_{\rm dip}$ couple the rate
equations at different excitation energies. That is the approach used
in Ref.~\cite{Wei11} for the perturbative regime ($\Gamma_{\rm dip}
\ll \hbar / \tau_{\rm coll}$) and in Refs.~\cite{Pal14, Pal15} in the
quasiadiabatic regime ($\Gamma_{\rm dip} \approx \hbar / \tau_{\rm
  coll}$). We stress that a rate equation may be used even for
excitation out of the ground state~\cite{Wei11}. For $N \approx 10^3 -
10^4$ that yields an excitation probability of order unity.

{\it Sudden Regime.} The results of the present paper cast new and
unexpected light on the (so far unexplored) sudden regime $\Gamma_{\rm
  dip} \gg \hbar / \tau_{\rm coll}$. In that regime photoabsorption is
so fast that nuclear relaxation may become negligible. But then the
use of rate equations is not justified. And we have shown above that
the classical approximation~(\ref{15}) does not yield significant
excitation because the electric field strength is too faint.
Significant nuclear excitation by laser-induced photon absorption is,
therefore, possible only when between any two absorption processes,
the nucleus does have time to relax. In other words, in the sudden
regime the characteristic time scale for photon absorption is set by
the collision time $\tau_{\rm coll}$, not by $\hbar / \Gamma_{\rm
  dip}$. That suggests that in the sudden regime, nuclear excitation
processes are only quantitatively but not qualitatively different from
the ones in the quasiadiabatic regime~\cite{Pal15}.

{\it Conclusions.} Within the dipole approximation, we have compared
laser-induced photon absorption processes in atoms and in nuclei,
assuming laser pulses of equal photon numbers but different mean
energies per photon ($0.5$ eV versus $5$ MeV) and  energy
spreads ($0.005$ eV versus $50$ keV).  If the lateral width of the laser is unchanged,
scaling shows that the electric field strength $| \vec{E}_{\rm cl} |$
scales  with the factor $10^7$ while the system radius scales with
a factor $10^{- 4}$. Therefore, the square of a typical dipole matrix
element of the interaction Hamiltonian (\ref{15}) scales with $| \vec{E}_{\rm cl}
|^2 R^2 \propto 10^6$ while the dipole transition rate (\ref{26}) scales with
$(\hbar \omega_0)^3 R^2 \propto 10^{13}$. That fact renders the
transition rate a much more important concept for the laser-nucleus
interaction than for the laser-atom interaction.

We have addressed two approximation schemes for quantum many-body
systems that allow for a practicable treatment of multi-photon
absorption processes. These are: (i) The mean-field approach, i.e.,
use of the time-dependent Schr{\"o}dinger equation for single-particle
motion, combined with the classical approximation~(\ref{15}) for the
electric field strength. This approach is characteristic for atoms, for which
the classical electric
field strength of the laser may be so strong as to significantly
deform the nuclear Coulomb potential, giving rise to 
ionization. We have shown that in nuclei, the scaled field strength is
too faint to cause significant excitation. (ii) 
 Rate equations that describe strongly interacting systems
which relax between any two subsequent photon absorption
processes. Rates are fundamentally a quantum concept. We have shown that rate
processes  yield significant nuclear excitation carrying the system
far above yrast~\cite{Pal15}.  Conversely, the transition rate is
relatively much less important in atoms. That fact coincides with the
smaller role of relaxation processes in atoms. The need to use rate
equations in nuclei casts new light on the case where the rate for
photon absorption is much bigger than the relaxation rate (the
``sudden'' regime). Here the nuclear relaxation rate (not the photon
absorption rate) should define the characteristic time scale for the entire
process.

The number $N \approx 10^4$ of photons required~\cite{Pal15} to
generate significant multiple photon absorption in nuclei via rate
processes is  much smaller than the number $10^{10}$ of photons
per pulse carried by a medium-intensity optical laser. Thus, coherent
Doppler backscattering of a tiny fraction of all photons in the
original atomic laser pulse suffices to generate significant nuclear
excitation. That fact reduces the requirements imposed on the
experimental realization of the ``flying mirror'' and strongly
enhances the likelihood of a successful start of experimental
laser-nucleus reaction processes.

The authors would like to thank P. Thirolf for fruitful discussions. This work is 
part of and supported by the DFG Collaborative Research Center ``SFB 1225 (ISOQUANT)''.

%%%%%%%%%%%%%%%%%%%%%%%%%%%%%%%%%%%%%%%%%%%%%%%%%%%%%%%%%%%%%%%%%%%%%%%%%
\bibliographystyle{apsrev}
\bibliography{sudden}
%%%%%%%%%%%%%%%%%%%%%%%%%%%%%%%%%%%%%%%%%%%%%%%%%%%%%%%%%%%%%%%%%%%%%%%%%%%%%%%%

\end{document}